\documentclass{ws-jai}
\usepackage[flushleft]{threeparttable}
\usepackage[dvipsnames]{xcolor}
\usepackage[hang,small,bf]{caption}
\usepackage[subrefformat=parens]{subcaption}
\usepackage[utf8]{inputenc}
\usepackage{ulem}
\usepackage{hyperref}
\usepackage{natbib}
\DeclareUnicodeCharacter{3000}{\ }
\begin{document}

\catchline{}{}{}{}{} 

\markboth{Hibiki Yama}{Optical Alignment Method for the PRIME Telescope}

\title{Optical Alignment Method for the PRIME Telescope}

\author{Hibiki Yama$^{1\dagger}$, Daisuke Suzuki$^{1}$, Shota Miyazaki$^{2}$, Andrew Rakich$^{3}$, Tsubasa Yamawaki$^{1}$, Rintaro  Kirikawa$^{1}$, \\Iona Kondo$^{1}$, Yuki Hirao$^{1}$, Naoki Koshimoto$^{4,5}$ and Takahiro Sumi$^{1}$}

\address{
$^{1}$Department of Earth and Space Science, Graduate School of Science, Osaka University, 1-1 Machikaneyama, Toyonaka, Osaka 560-0043, Japan; yama@iral.ess.sci.osaka-u.ac.jp, \\
$^{2}$Institute of Space and Astronautical Science, Japan Aerospace Exploration Agency, 3-1-1 Yoshinodai, Chuo, Sagamihara, Kanagawa 252-5210, Japan\\
$^{3}$Mersenne Optical Consulting, New Zealand\\
$^{4}$Laboratory for Exoplanets and Stellar Astrophycis, NASA Goddard Space Flight Center, Greenbelt, MD 20771, USA\\
$^{5}$Department of Astronomy, University of Maryland, College Park, MD 20742, USA}

\maketitle

\corres{$^{\dagger}$Corresponding author}

\begin{history}
\received{(to be inserted by publisher)};
\revised{(to be inserted by publisher)};
\accepted{(to be inserted by publisher)};
\end{history}

\begin{abstract}
We describe the optical alignment method for the 
Prime-focus Infrared Microlensing Experiment (PRIME) telescope 
which is a prime-focus near-infrared (NIR) telescope with a wide field of view for the microlensing planet survey toward the Galactic center that is the major task for the PRIME project.
There are three steps for the optical alignment: preliminary alignment by a laser tracker, fine alignment by intra- and extra-focal (IFEF) image analysis technique, and complementary and fine alignment by the Hartmann test. We demonstrated that the first two steps work well by the test conducted in the laboratory in Japan.
The telescope was installed at the Sutherland Observatory of 
South African Astronomical Observatory in August, 2022. 
At the final stage of the installation, we demonstrated that the third method works well and the optical system satisfies the operational requirement. 
\end{abstract}

\keywords{the PRIME telescope; optical alignment; Hartmann test.}

\section{Introduction}
\label{sec_intro}

The PRime-focus Infrared Microlensing Experiment (PRIME) telescope is a prime-focus near-infrared (NIR) telescope with a wide field of view (FOV) to conduct the microlensing planet survey toward the Galactic center \cite{Kondo23}.
So far, microlensing planet surveys have been conducted by using the dedicated optical telescopes operated by the survey teams, such as, MOA \citet{bon01, sum03}, OGLE \cite{uda15}, and KMT-Net \cite{kim16}, because such surveys require large number of pixels, which can be relatively easily realized by CCDs compared to NIR detectors.
NIR bulge surveys also have been conducted by VISTA \cite{nav17, nav18, nav20a} and UKIRT \citet{shv17, shv18}, but the cadence of these survey data is not high enough to detect short time transients such as planetary anomaly signals in the microlensing light curves.
PRIME will conduct such a high cadence imaging survey toward the Galactic center and bulge with NIR for the first time. 
NIR can mitigate the dust extinction so that we can monitor the higher stellar density fields such as the Galactic plane and center, which are not accessible with the ordinary optical microlensing surveys.
The Galactic bulge survey by the PRIME telescope is expected to put further constraint on the cool planet mass function of bound \cite{suz16} and unbound planets \cite{sum11, mro17, gou22, kos23, sum23}, to optimize the Roman Galactic Bulge Time Domain Survey field \cite{pen19, joh20} by providing the microlensing event rate toward the inner bulge, to find the isolated black hole candidates \cite{lam22, sah22}, to study variable stars in the bulge \cite{sos19}, to give a new constraint on the structure of the galactic center \cite{kos21} and so on. 
In addition, when the bulge is below the horizon, the PRIME telescope will observe transient objects \cite{dur22}, such as gravitational wave events, super novae, transiting planets, etc.

To realize such scientific observations, the telescope should satisfy the required optical design.
In this paper, we describe how the optical alignment of the PRIME telescope was conducted. 
There are three steps for the optical alignment: 
preliminary alignment by a laser tracker, 
fine alignment by intra-focal and extra-focal (IFEF) image analysis technique, and 
complementary and fine alignment by the Hartmann test.
Because the construction of the main instrument, PRIME-Cam, was delayed compared to that of the PRIME telescope, the optical alignments were conducted before the installation of PRIME-Cam.

Designs of the PRIME telescope and PRIME-Cam are introduced in Section \ref{sec_tel}.
Three optical methods are summarized in Section \ref{sec_method}.
The first step of the optical alignment is written in Subsection \ref{sec_faro}. A fine alignment by using the IFEF images as the second step is described in Subsection \ref{sec_IFEF}. The third step by the Hartmann test is written in Subsection \ref{sec_hartmann}. We used these methods at the final phase of the telescope construction and the result is summarized in Section \ref{sec_res}. Finally, we conclude our method in Section \ref{sec_con}.

\section{The PRIME Telescope}
\label{sec_tel}

\begin{figure}
    \centering
    \includegraphics[width=100mm]{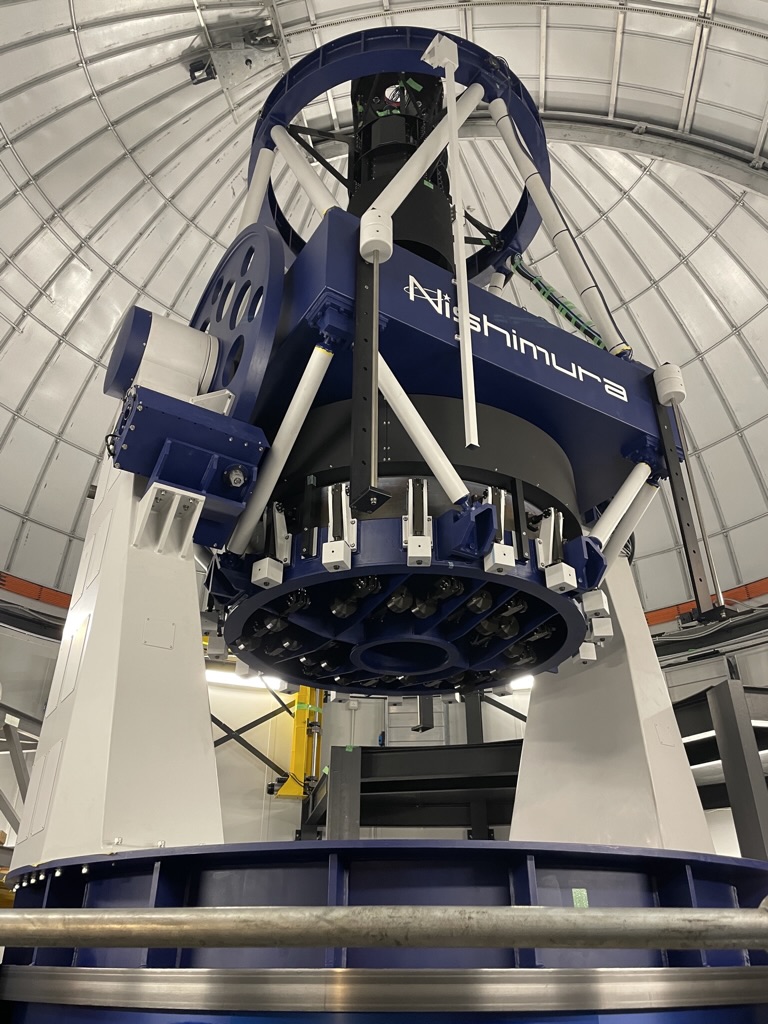}
    \caption{The PRIME telescope at Sutherland observatory, SAAO in August, 2022.}
    \label{fig:primetel}
\end{figure}

\begin{figure}
    \centering
    \includegraphics[width=140mm]{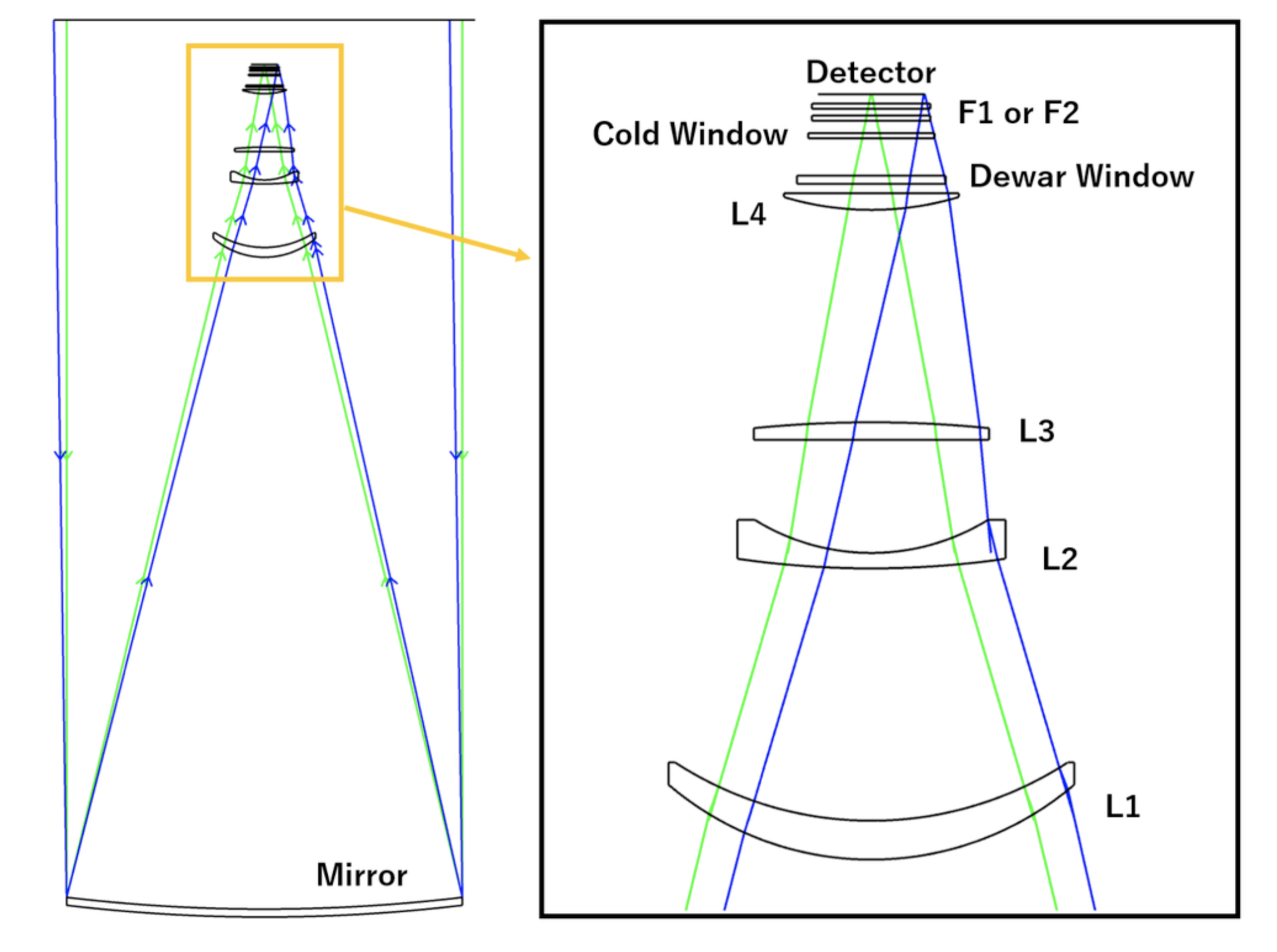}
    \caption{The optical design of the PRIME telescope and PRIME-Cam. The Prime Focus Unit (PFU)　consists of L1, L2, L3 and L4. The optical alignments are conducted by adjustment of PFU decenter and tip/tilt. In addition, only L2 is movable, so we can do fine adjustments with L2 tip/tilt. Note that the instrument is fixed to PFU.}
    \label{fig:primeoptdesign}
\end{figure}

The PRIME telescope is a prime-focus near-infrared telescope. It consists of a 1.8m diameter, parabolic primary mirror and four corrector lenses that are held in the Prime Focus Unit (PFU).
Figure \ref{fig:primetel} shows the appearance of the telescope.
The primary objective is to conduct a NIR microlensing planet survey, which requires a high throughput in the NIR wavelengths and a wide field of view.
An imaging instrument, PRIME-Cam is mounted on the primary focus. In the near future, a flip mirror for a spectrograph\footnote{South Africa Near-infrared Doppler (SAND)} will be installed between the second (L2) and third lenses (L3).
Figure \ref{fig:primeoptdesign} shows the optical design of the telescope and the specifications of the telescope is summarized in Table \ref{tab:primedetail}.
The telescope was installed at Sutherland Observatory, South African Astronomical Observatory (SAAO) in August, 2022.

\begin{table}
\caption{Specification of the PRIME telescope.}
    \centering
    \begin{threeparttable}
    \begin{tabular}{l||c} \hline
         & PRIME telescope  \\ \hline \hline
       Primary Mirror & $1.8$\,m,\,parabolic  \\ 
       F-number & 2.29 \\ 
       Prime Focus Unit & modified Wynne-type system\tnote{a} \,(four lenses) \\
       Focus position & Prime focus \\
       site & Sutherland Observatory \,(South Africa) \\ \hline
    \end{tabular}
    \begin{tablenotes}
    \item[a]
    \small{\cite{Raki13}}
    \end{tablenotes}
    \end{threeparttable}
    \label{tab:primedetail}
\end{table}

\subsection{PRIME-Cam}
\label{sec_prime_cam}
PRIME-Cam is the primary imaging instrument of the PRIME telescope. 
It is a wide FOV NIR camera that was mounted on the prime focus of the telescope in October 2022. 
The pixel scale of PRIME-Cam is $0.513"/\rm pix$ and FOV is $1.45\, \rm{deg}^2$ including the gaps between the detectors. This instrument was made at NASA Goddard Space Flight Center.
It uses four Teledyne HAWAII 4RG (H4RG-10) detectors, each of which is a $4 \rm{K} \times 4\rm{K}$ HgCdTe photodiode arrays with a $10\,\mu\rm{m}$ unit cell size \citep{mos20}. 
The detectors are cooled down to $\sim100\,\rm{K}$ by two cryocoolers.

\begin{figure}
    \centering
    \includegraphics{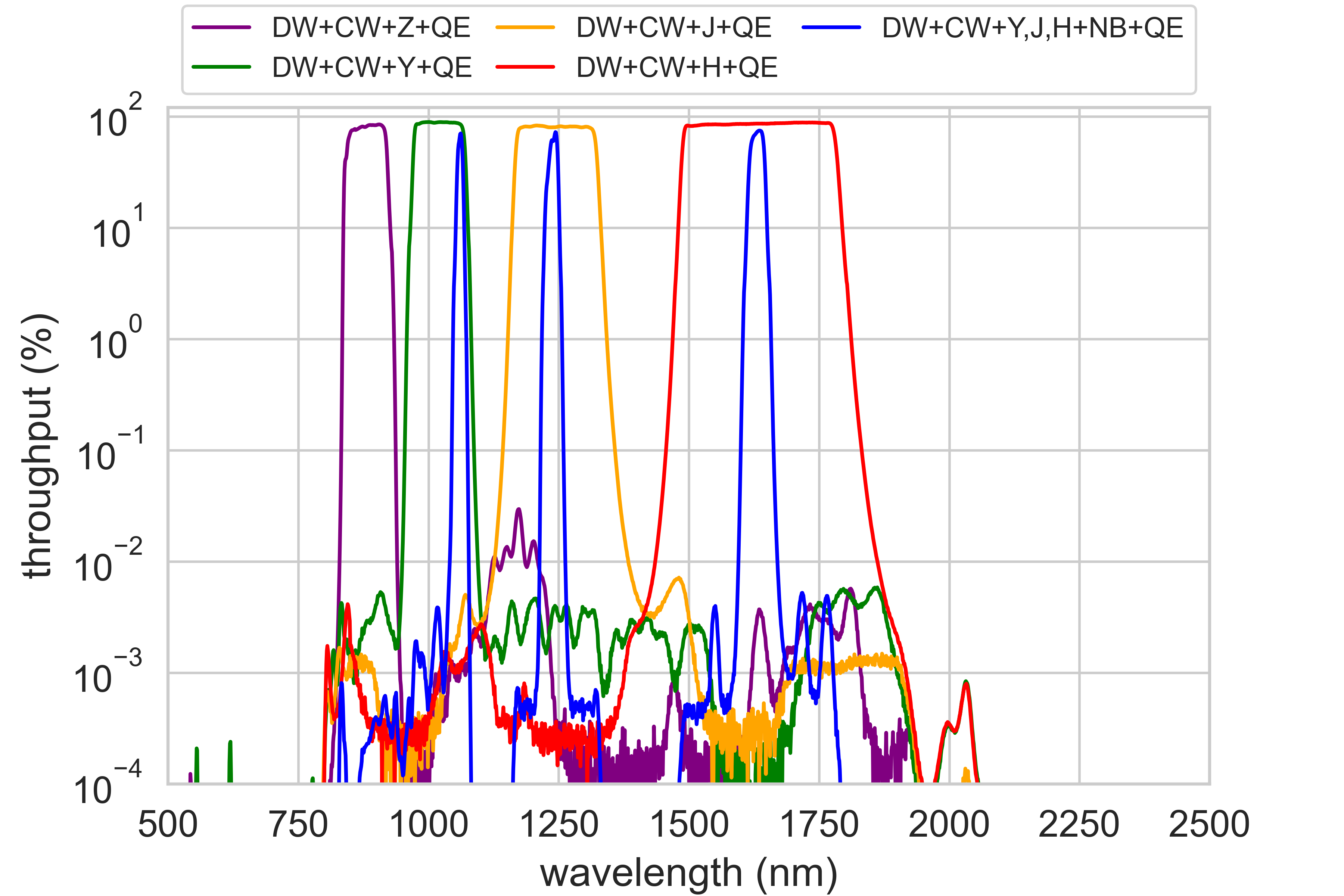}
    \caption{Throughput of PRIME-Cam. This throughput is the transmittance of all optical elements of PRIME-Cam multiplied by the QE of the H4RG-10.}
    \label{fig:prime_filters}
\end{figure}

Because the detectors and filters should be cooled down to decrease dark current and thermal radiation, these are inside the dewar container which is vacuumed at $\sim 10^{-6} \,\rm{mbar}$. 
There are optical elements on the top of and inside the dewar; a dewar window (DW), a cold window (CW), two filter wheels with five filters (Wheel1 : $Z$-band filter, $\it{3} Narrow$-band filter, $Dark$, $Open$, Wheel2 : $Y$-band filter, $J$-band filter, $H$-band filter, $Open$) and four detectors. 
The three narrow-band filter ($\it NB1063$, $\it NB1243$, and $\it NB1630$) is used with $Y$-, $J$- or $H$-band filter to pick up one narrow bandwidth.
The cold window made of BK7 is designed to cut a shorter wavelength than the $Z$-band and a longer wavelength than $H$-band. All filters are made of BK7 and their throughput is plotted in Figure \ref{fig:prime_filters}.  All of these PRIME-Cam optical elements have no curvature and are parallel plates. 
The optical elements of the PRIME-Cam are plotted in Figure \ref{fig:primeoptdesign}.
The details of the PRIME-Cam is summarized in Table \ref{tab:primecamdetail}.

\begin{table}
\caption{Specification of the PRIME-Cam.}
    \centering
    \begin{threeparttable}
    \begin{tabular}{l||c} \hline
         & PRIME-Cam  \\ \hline \hline
         Field of view (FOV)\tnote{a} & $1.45\,\mathrm{deg}^2$　\\
         Detectors & 4\,$\times$\,H4RG-10\,\,(HgCdTe) \\ 
         Temperature of detector & $\sim100\,\rm{K}$ \\
         Total pixel & 67\,M\,pixel \\
         Pixel size & 10\,$\rm{\mu m}$ \\
         Pixel scale & 0.513\,arcseconds \\
         Filters & 
       \begin{tabular}{r}
            $Z$,\,$Narrow$-band \,(wheel1) \\
            $Y$,\,$J$,\,$H$ \,(wheel2) 
       \end{tabular} \\ \hline
    \end{tabular}
    \begin{tablenotes}
    \item[a]
    \small{Including a gap between detectors.}
    \end{tablenotes}
    \end{threeparttable}
    \label{tab:primecamdetail}
\end{table}

\section{Optical Alignment Methods}\label{sec_method}
\begin{figure}
  \begin{center}
    \begin{tabular}{c}
      \begin{minipage}{0.5\hsize}
        \begin{center}
          \includegraphics[clip,width=9.0cm]{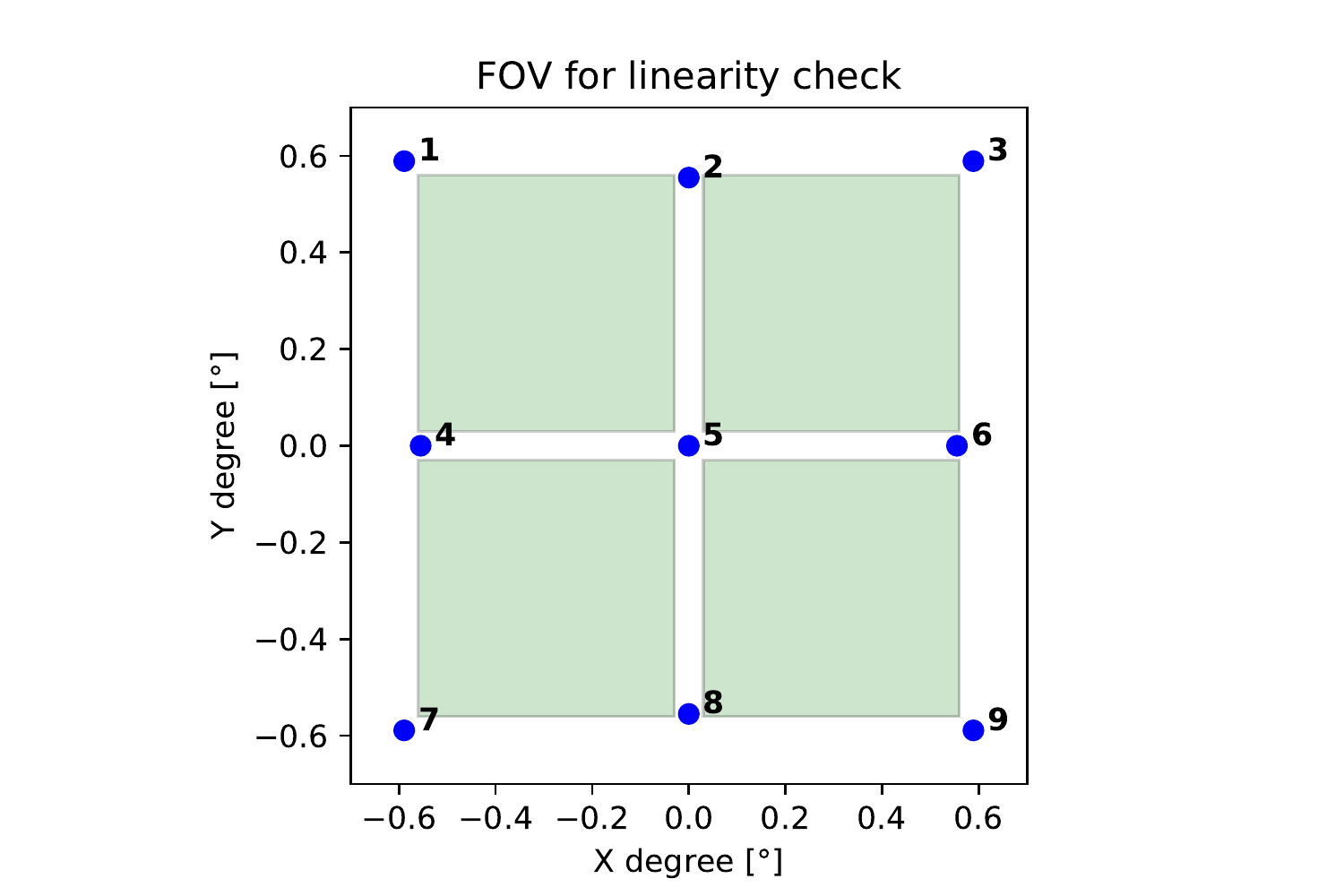}
          \subcaption{Blue dots are positions where the simulation was conducted. Green areas are detectors.}
          \hspace{1.6cm} 
        \end{center}
      \end{minipage}
      \begin{minipage}{0.5\hsize}
        \begin{center}
          \includegraphics[clip,width=6.5cm]{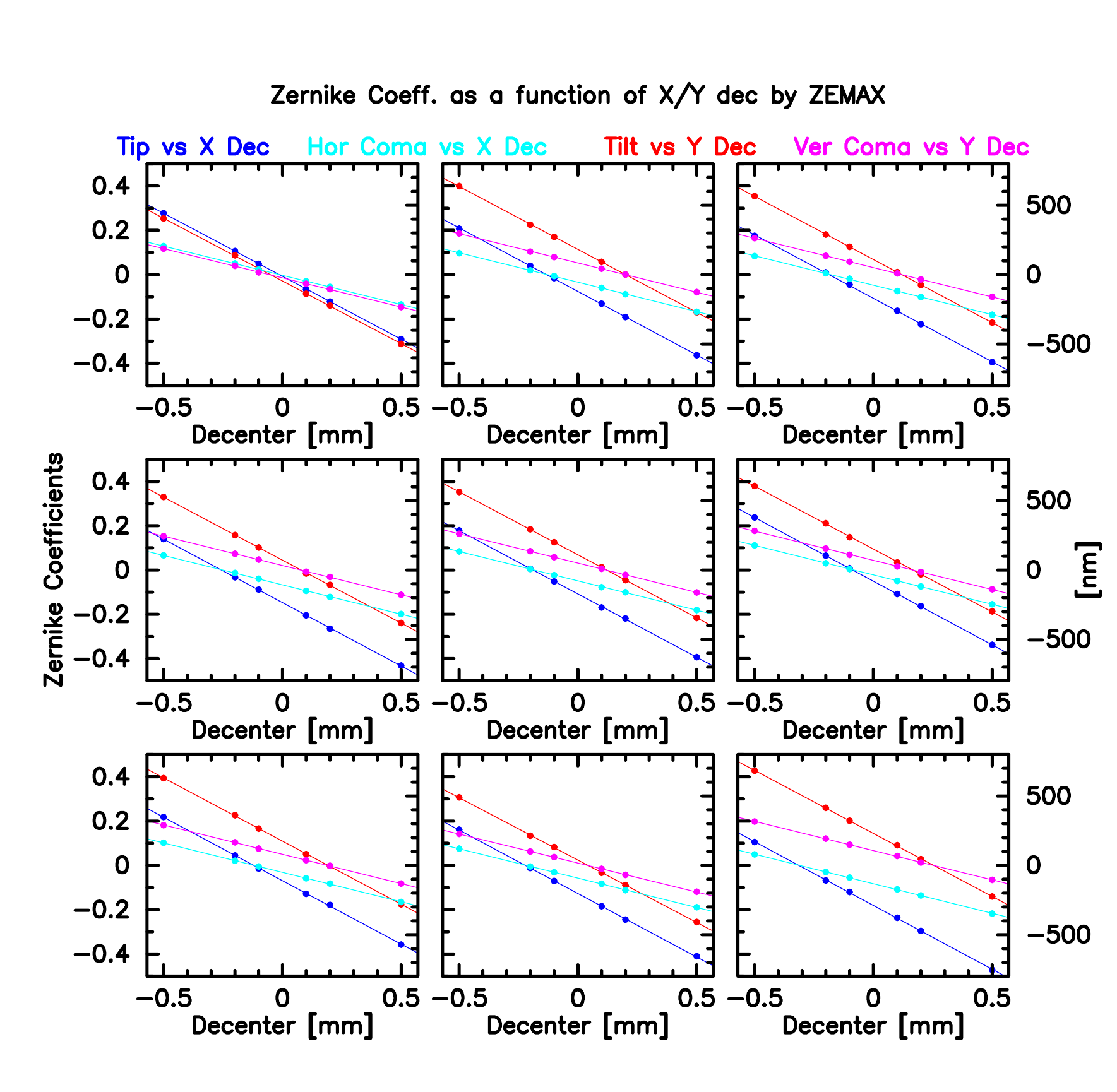}
          \subcaption{Tip/Tilt and Coma which are equivalent to PFU Decenter. }
          \hspace{1.6cm} 
        \end{center}
      \end{minipage}
      \\
      \\
      \begin{minipage}{0.5\hsize}
        \begin{center}
          \includegraphics[clip,width=6.2cm]{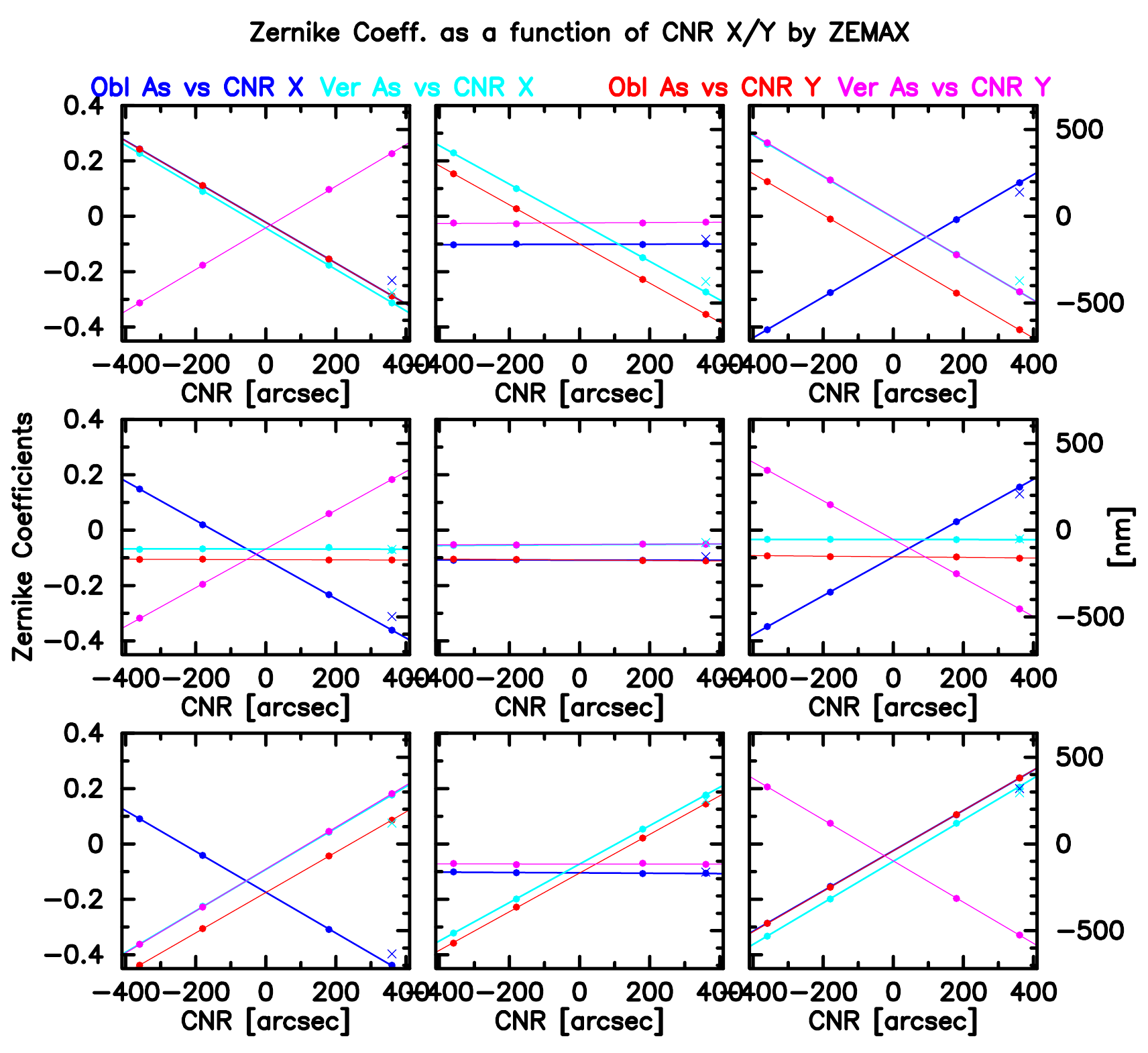}
          \subcaption{Astigmatisms which are equivalent to PFU CNR.}
          \hspace{1.6cm} 
        \end{center}
      \end{minipage} 
      \begin{minipage}{0.5\hsize}
        \begin{center}
          \includegraphics[clip,width=6.5cm]{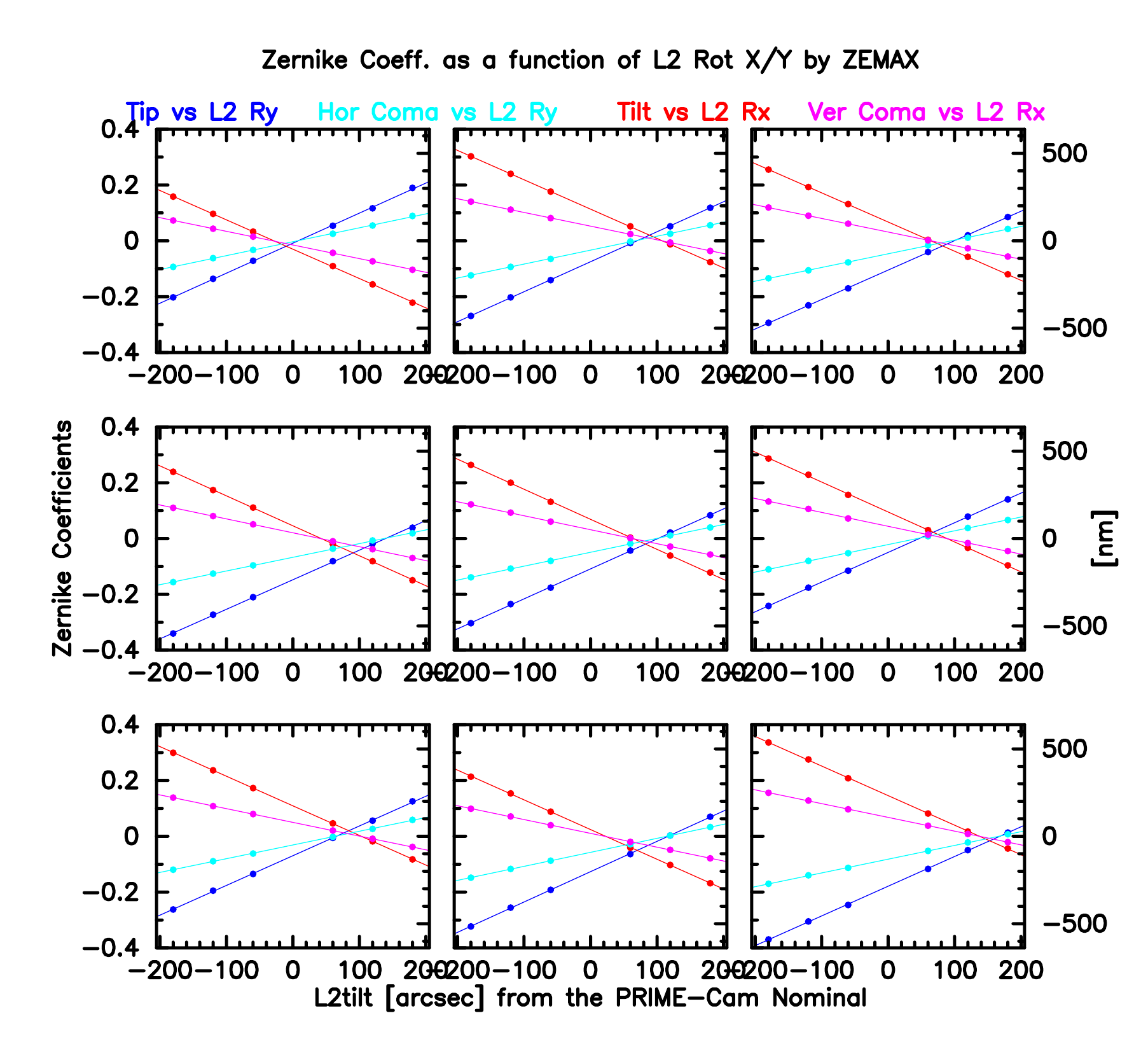}
          \subcaption{Tip/Tilt and Coma which are equivalent to L2 tip/tilt.}
          \hspace{1.6cm} 
        \end{center}
      \end{minipage}
    \end{tabular}
    \caption{The simulation of the major aberrations which are equivalent to the misalignment of the PFU and the L2. (a) We evaluated aberrations at the nine different positions (blue dots) within the FOV of the PRIME-Cam by using the simulation images of the Hartmann test. (b)-(d) We simulated the Zernike coefficient of major aberrations by PFU decenter, PFU CNR and L2 tip/tilt. The nine tiles in each panel correspond to the nine different positions within the FOV of the PRIME-Cam. } 
    \label{fig:LinearitySimulations}
  \end{center}
\end{figure}

We should align the telescope optics within tolerances to achieve maximum scientific results.
The major aberrations we can correct in the optical alignment are coma and astigmatism, the amount of which depends on the decenter and coma neutral rotation (CNR) of the PFU linearly.
The CNR is the tip/tilt of the PFU around the coma neutral point that exists on the optical axis. The CNR does not affect the coma so we can independently correct the coma and astigmatism. 
We cannot rotate the PFU around the coma neutral point, but the CNR can be realized by the combination of the decenter and PFU tip/tilt.
Also, the second largest lens, L2, is adjustable for piston and tip/tilt by three actuators. The L2 tip/tilt can correct the field linear coma (there is some correlation to other aberrations though). Hence, we can correct the coma and astigmatism by PFU decenter and PFU CNR, respectively. Basically, L2 is fixed to the reference position at the manufactured, and L2 is moved when we cannot complete alignment by adjusting the PFU decenter and PFU CNR.
Figure \ref{fig:LinearitySimulations} shows the simulation of the major aberrations which are equivalent to the misalignment of the PFU and the L2.

The operational tolerance is derived based on the optical simulation. The PFU tip/tilt and decenter should satisfy $<22"$ and $<280\,\mu\rm{m}$, which are equivalent to $80\%\,\mathrm{Encircled\,Energy\,radius}< 0.40"$ based on the $15\%$ degradation of the nominal design\footnote{The PRIME telescope was designed to meet $80\%\,\mathrm{Encircled\,Energy\,radius}< 0.35"$.}.
We simulated the Hartmann constant of the optical system by adding tip/tilt and decenter of the PFU within the operational tolerance and found that the worst Hartmann constant within the tolerance is 0.314" at $H$-band for the average over the inner FOV. Note that typical seeing at $H$-band at Sutherland observatory is $\sim$ 1.4". To satisfy the operational tolerance, we conduct three steps of the optical alignment. Below we describe each of the three methods.

\subsection{Preliminary Alignment by Laser Tracker}
\label{sec_faro}
We first conduct the alignment of the optical axis between the primary mirror and PFU using a commercial laser tracker, FARO\footnote{https://www.faro.com/ja-JP/Products/Hardware/Vantage-Laser-Trackers}.
By conducting this procedure, we can achieve the axis alignment of tip/tilt and decenter within several tens of arcseconds and a hundred $\mu$m precision, respectively.
We also emphasize that this takes only a few hours, which is much faster than an optical alignment using star images for achieving such precision.

As a preparation for the axis alignment, we measured the 3D positions of the optical axes of the primary mirror and PFU.
For the primary mirror, we measured the surface of the mirror using the FARO laser tracker and fitted the data points assuming that the mirror shape is a paraboloid, and then estimated the optical axis of the primary mirror.
As for PFU, the lens manufacturer company reported the measurements of the optical axis and mechanical reference planes in PFU.
We measured the surfaces of the reference planes using the FARO laser tracker and then estimated the optical axis of PFU. Note that systematic uncertainty is in the models to estimate the optical axis of PFU and the primary mirror.
After that, we attached several reference points that can mount Spherically Mounted Retroreflector (SMR) in the mirror and PFU respectively, and then measured the positions of the reference points.
By doing this, we can get the relative positions of the optical axis from the reference points, in other words, we can immediately obtain the 3D positions of the optical axis by measuring the reference points on both the primary mirror and PFU using the FARO laser tracker.
Due to this measurement, we can conduct the alignment of the optical axis between the primary mirror and PFU.




\subsection{Fine Alignment in $z$-band by IFEF Image Technique}
\label{sec_IFEF}

We can correct the major aberrations from the optics by using image analysis of two defocused pupil images \cite{kui04, kuirak14}.
These images are obtained with two defocused CMOS detectors, symmetrically displaced in both radial and axial directions from a central, ``on-axis" and ``in-focus" detector.
One of the displaced detectors is closer to the primary mirror than the in-focus detector, or ``intra-focal" (IF) and the other detector is further from the primary mirror than the in-focus detector, or ``extra-focal" (EF).

\begin{figure}
\begin{center}
\includegraphics[width=100mm]{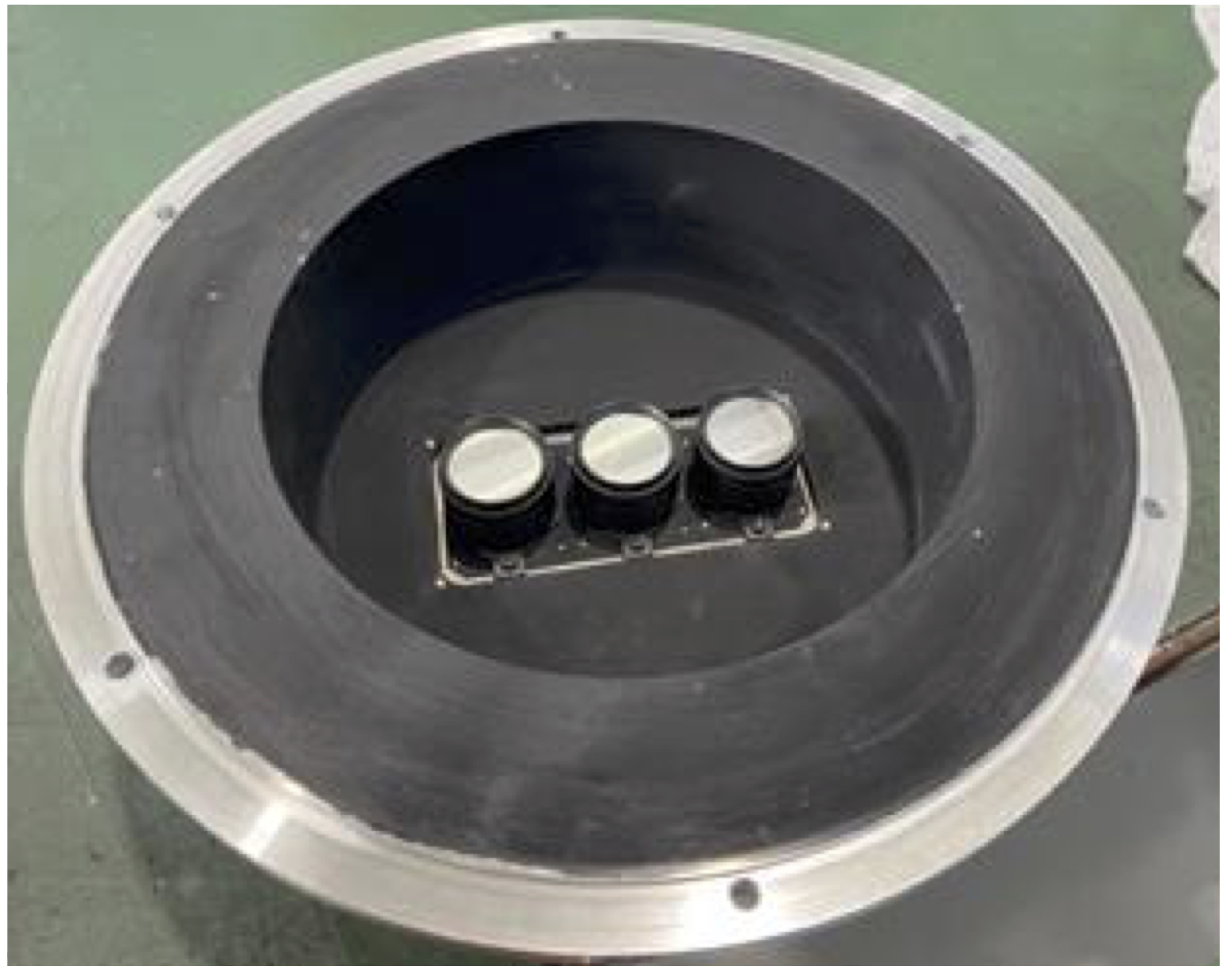} 
\end{center}
\caption{Picture of $z$-band test camera that has three CMOS detectors mounted on the center and edges of the FOV, respectively.}
\label{fig:yamawaki-c}
\end{figure}

\begin{figure}
\begin{center}
\includegraphics[width=100mm]{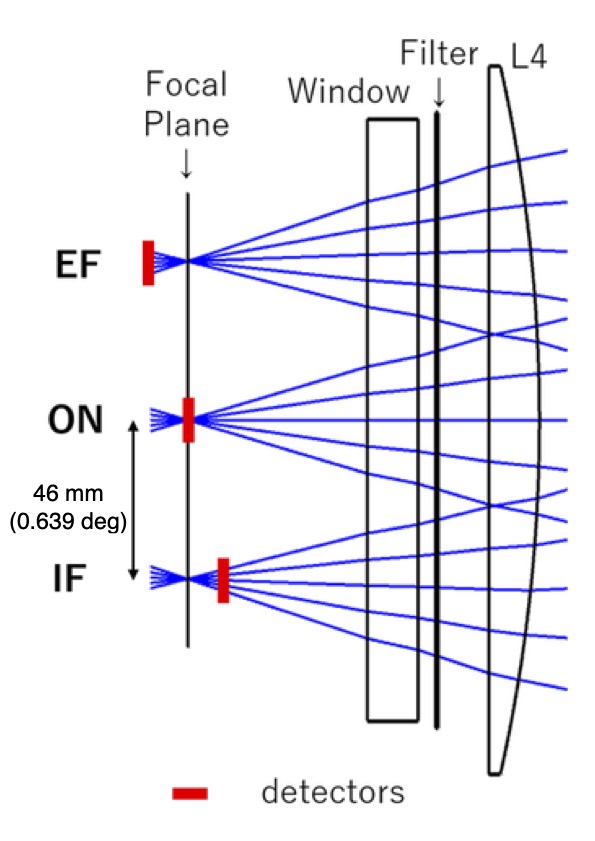} 
\end{center}
\caption{Schematic view of the optical design of the $z$-band test camera. The 46mm offset between ON camera and IF (EF) camera corresponds to $0.639$ deg.}
\label{fig:opt_design_yamawaki-c}
\end{figure}

We designed and built a camera dedicated to this purpose by using three CMOS sensors. 
Figure \ref{fig:yamawaki-c} and Figure \ref{fig:opt_design_yamawaki-c} are a visual appearance of the camera and a schematic view of optical design, respectively.
The IF and EF cameras are off-axis by 46mm and defocused by 0.7mm from the ON camera which is on-axis and in-focus. 
CMOS sensors can be obtained at a reasonable cost, whereas they are not sensitive to NIR wavelengths such as $H$-band.
As for the sensors, we use three ARTCAM-092XQE-WOM CMOS cameras.
We use the $z$-band ($\sim 900\, \rm nm$), as its wavelength is the longest (closest to the $H$-band) among the wavelengths to which the sensors are sensitive enough.

\begin{table}[]
    \centering
    \caption{Type of the Aberration Source and its Amount for the Basis Function}
    
    \begin{tabular}{rcc}
    \hline \hline
    Aberrations Source  & Amount of Changes  &  Unit\\ \hline
    Corrector piston & [-200, 200] & $\mu$m \\
    L2 piston & [-200, 200] & $\mu$m \\
    L2 tilt Rx & [-200, 200] & arcsec \\
    L2 tilt Ry & [-180, 180] & arcsec \\
    CNR tilt Rx & [-180, 180] & arcsec \\
    CNR tilt Ry & [-180, 180] & arcsec \\
    M1-Z5 & [-400, 400] & nm \\
    M1-Z6 & [-400, 400] & nm\\
    M1-Z11 & [-200, 200] & nm\\
    Decenter X & [-500, 500] & $\mu$m\\
    Decenter Y & [-500, 500] & $\mu$m\\
    Linear coma Z7 & [-400, 400] & nm\\
    Linear coma Z8 & [-400, 400] & nm\\
    \hline
    \end{tabular}
    
    \begin{tablenotes}
    \item[] \small{ 
    \textbf{Note.} Z5 and Z6 are field constant, oblique and vertical astigmatism. Z7, Z8 and Z11 are vertical coma, horizontal Coma and spherical aberration, respectively.}
    \end{tablenotes}

    \label{table:aberrations}

\end{table}

To analyze the defocused images, we consider low-order aberrations which are supposed to be dominant ones we can measure.
The aberrations we use for the modeling are summarized in Table \ref{table:aberrations}.
The ZEMAX model we use for this simulation includes the measured lens positions, but it does not include the surface information of the lenses and mirror.
It is reported that L1 (the largest lens) has an issue with the polishing on the convex surface, which would induce some amount of field constant astigmatism. 
The primary mirror would also give some constant astigmatism, but this would be in a different direction and depend on the telescope elevation angle.
The optical design itself has residuals of low-order aberration including spherical aberration on- and off-axis and coma and astigmatism off-axis, and these need to be taken into account during alignment.
Therefore, we model these effect as M1-Z5, -Z6, -Z11, Linear coma Z7, and Linear coma Z8 as indicated in Table \ref{table:aberrations}. 
Note that we use the Zernike standard coefficients.

\begin{figure}
    \centering
    \includegraphics[width=160mm]{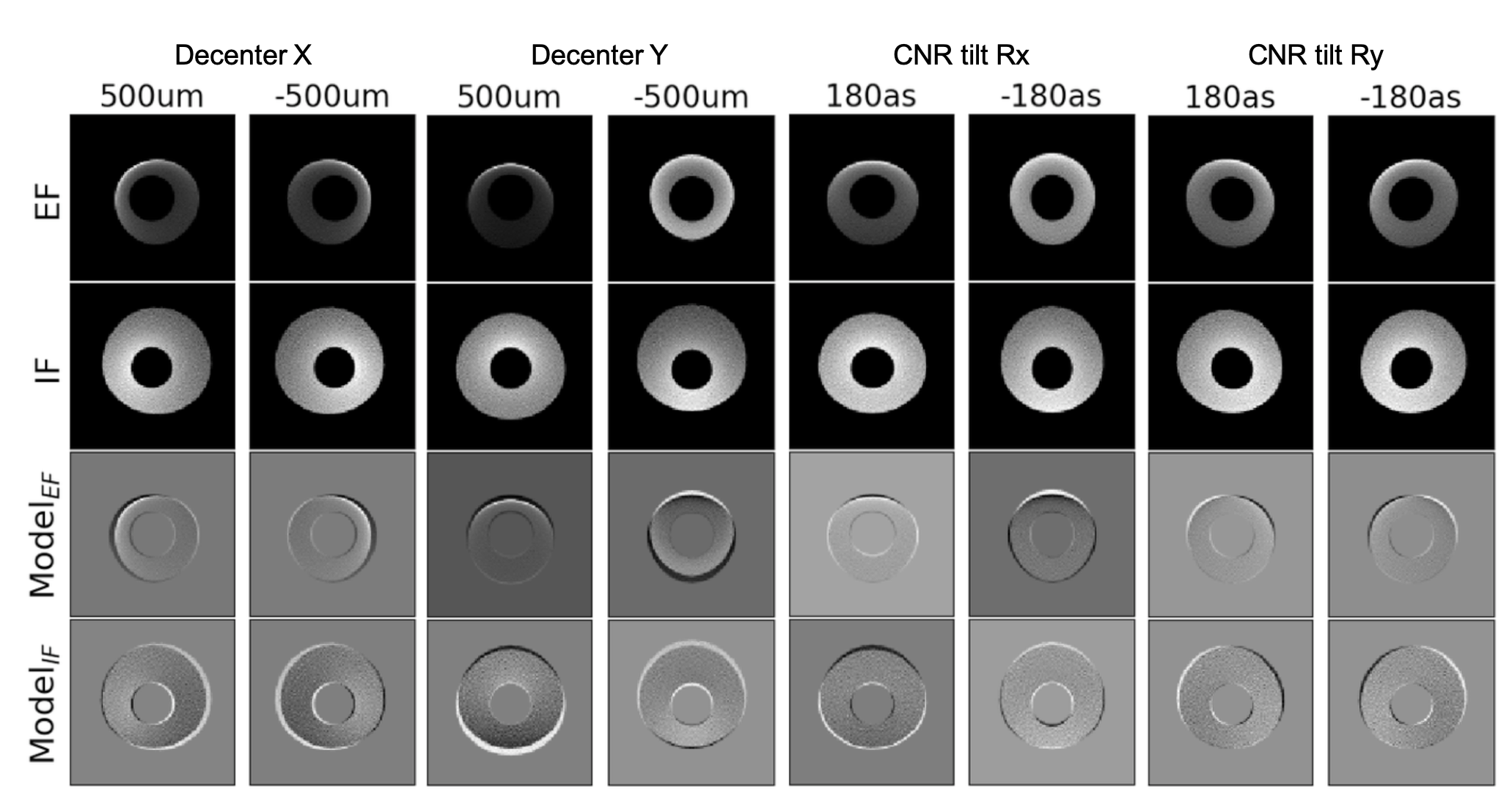}
    \caption{Simulated defocus images for the decenter and CNR tilt. The bottom two panels show the simulated images where the nominal images are subtracted from the IF/EF images.}
    \label{fig:defocusedmodel}
\end{figure}

Following the method of \citet{kui04}, we generate basis functions, $Model_{i}$. The observed defocused images can be explained as a combination of the basis functions,
\begin{align}
    Image_{\rm model} &= K * (Image_{\rm nominal} + \sum_{i}{A_{i}Model_{i}}) \\
    Model_{i} &= Image_{i} - Image_{\rm nominal}    
\end{align}
where $Image_{\rm model}$ is a modeled defocused image, $Image_{\rm nominal}$ is the nominal defocused image without aberrations, $Image_{i}$ is a simulated image with the $i$th aberration, $K$ is the seeing and $A_i$ is the fitting parameter to minimize the difference between an observed defocused image and $Image_{\rm model}$.
Some examples for the modeled defocused images are shown in Figure \ref{fig:defocusedmodel}.

We evaluated the sensitivity of this method to PFU decenter, PFU CNR and L2 tip/tilt by using some simulation images under the typical seeing condition. The sensitivity to PFU decenter　is within a several tens of $\mu$m and PFU CNR and L2 tip/tilt are both within a few tens of as.

Note that this method uses IF and EF images that are observed with the ``off-axis'' cameras, with which we can put constraint on the astigmatism. We cannot measure the astigmatism from the misalignment of the optics with the on-axis camera.
Because the PRIME telescope has a wide FOV, mitigating astigmatism and making the PSF shape as uniform as we can over the whole FOV is critical for the image quality.

\subsection{Fine Alignment in $H$-band by Hartmann Test}
\label{sec_hartmann}

The alignment method by the IFEF images described in Subsection \ref{sec_IFEF} uses a $z$-band filter as the detectors are CMOS sensors, whereas the microlensing survey will primarily use an $H$-band filter with H4RG detectors to mitigate the dust extinction toward the Galactic center.
Since the telescope design is optimized in $H$-band and there are some chromatic aberrations, we should conduct fine optical alignment using $H$-band. We will use a commercial InGaAs camera for this purpose as described below.
Such a camera is also useful to evaluate the performance of the telescope such as total throughput and limiting magnitude before the installation of the PRIME-Cam.

We use a classical Hartmann test to conduct the fine optical alignment method with $H$-band. Usually, the Hartmann test is used for the assessment of optics by measuring the Hartmann constant, which is the average distance from the barycenter of rays to each ray at the focal plane. One might qualitatively see the wavefront aberrations from the distribution of the Hartman spots. Here we developed software that can quantitatively extract the wavefront aberrations from the images of the Hartmann test.
Below, we briefly describe the $H$-band test camera. Then, the fine optical alignment method with the Hartmann test is written.

\subsubsection{H-band Test Camera (Yama-Cam)}
We developed an $H$-band test camera (Yama-Cam) using an IMX990 chip, that is, a $\rm{1K}\times\rm{1K}$ InGaAs detector by Sony. We can use this camera for the alignment described in this chapter, and conduct test observation before the installation of PRIME-Cam.

The IMX990 is sensitive to NIR wavelengths up to $\sim 1700\, \rm nm$ including the shorter half of the $H$-band and its unit cell size is $5\,\rm{\mu m}$ $\times$ $5\,\rm{\mu m}$, smaller than the size of PRIME-Cam detectors.
This is the reason why we chose this detector for the $H$-band test camera. This chip is usually chilled to $15^\circ$C with a Peltier device. However, we changed the setting of the Peltier device to make the temperature of the chip down to $3^\circ$C (the lower limit of temperature stability), because we want to decrease dark current for astronomical use. 
This chip is placed at almost the same distance from the L4 as the PRIME-Cam detectors. 
We measured gain, full well capacity, readout noise and dark current of the IMX990 chip we use for the $H$-band test camera and summarized them in Table \ref{table:IMX990} with the catalog values. 


\begin{table}[]
\centering
\begin{threeparttable}
  \caption{Specification of the IMX990}
  \begin{tabular}{lcc}
      \hline \hline
      & Catalog & Measurement\\ \hline 
      Image size &  $6.48\,\rm mm$(H) $\times$ $5.16\,\rm mm$(V) & --\\ 
      Effective pixel & $1296 \rm{(H)} \times 1032 \rm{(V)}$ & -- \\ 
      Unit cell size & $5\,\rm{\mu m}$ $\times$ $5\,\rm{\mu m}$ & -- \\
      Quantum efficiency (at $1.2\mu\rm{m}$) & $>\,75\%$ & --\\
      Gain[12bit] ($e^-/\mathrm{ADU}$) & -- & $26.9\,\pm 0.6$ \\
      Full well capacity ($10^3 e^-/\mathrm{pix}$) & $120$ & $110\, \pm \, 2$ \\
      Readout noise ($e^-/\mathrm{pix}$) & $<\,200$ & $142\,\pm 4$  \\
      Dark current\tnote{a} \, ($e^-/\mathrm{pix}/s$) & $850$ & $458$  \\
      \hline
  \end{tabular}
  \begin{tablenotes}
  \item[a] 
  \small{Typical values at $15^\circ \rm{C}$ and $3^\circ\rm{C}$ for the catalog and measurement, respectively.}
  \end{tablenotes}

 \label{table:IMX990}
\end{threeparttable}
\end{table}
\vspace{3mm}

\begin{figure}
    \centering
    \includegraphics{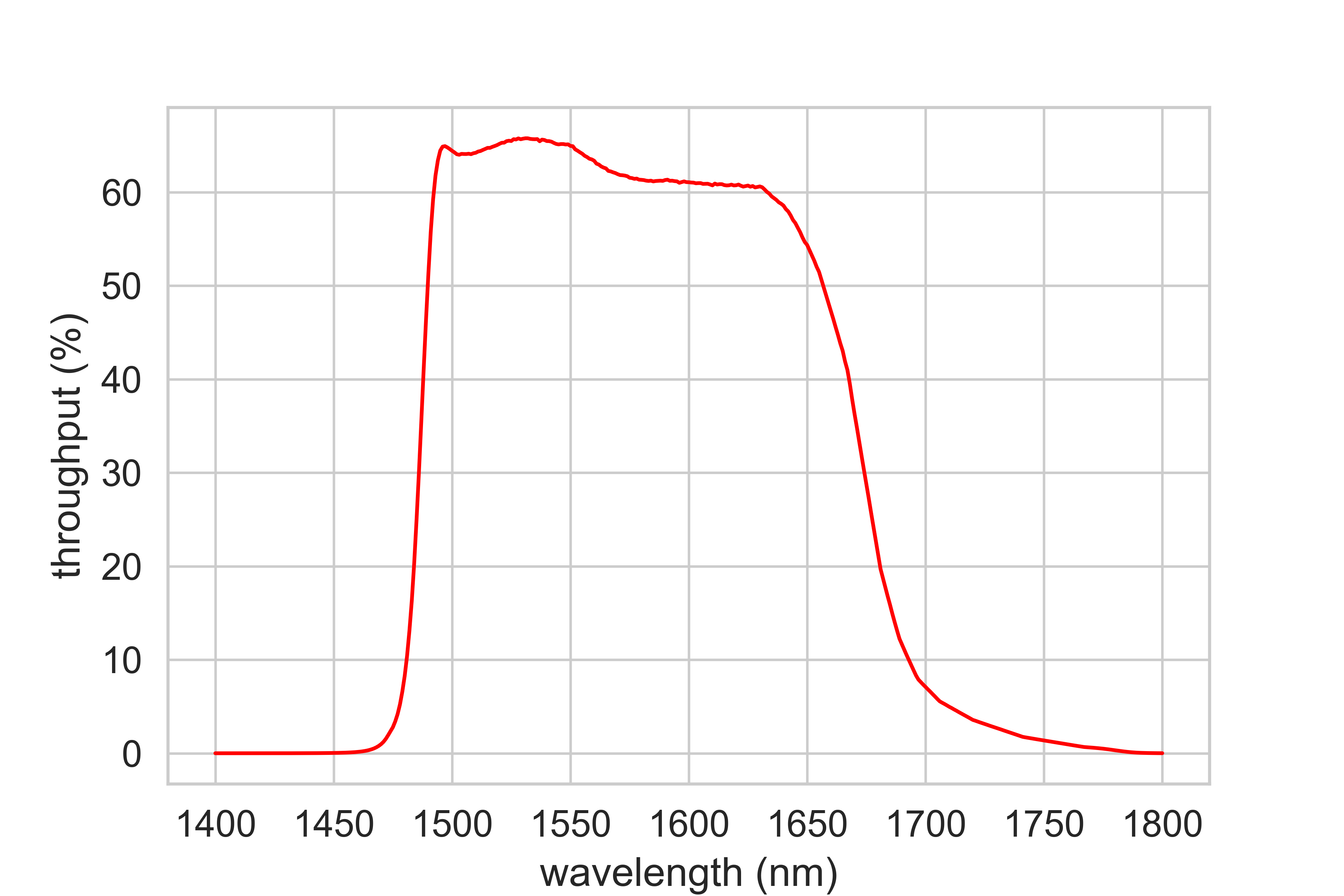}
    \caption{$H$-band throughput of Yama-Cam. This throughput is the transmittance of all optical elements  of Yama-Cam multiplied by the QE of the IMX990.}
    \label{fig:yama_cama_trans}
\end{figure}

Yama-Cam has two main optical elements. The one is a BK7 glass plate with Anti-Reflective coating in the range of wavelengths from $1500\,\rm{nm}$ to $1700\,\rm{nm}$ (Window). 
The other is an $H$-band pass filter which was manufactured in the same sputtering process as a filter for PRIME-Cam. This filter can be changed to other filters ($Z$, $Y$ and $J$, which were also made in the same sputtering process as the  PRIME-Cam filters). 
The detector has a small optical element: Cover glass. The cover glass is made of borosilicate glass. This element is needed for protecting the surface of the detector and is sealed with nitrogen to prevent condensation on the detector.
Figure \ref{fig:yama_cama_trans} shows the $H$-band throughput of the Yama-Cam.
The differences in the material are that the window, the $H$-band pass filter and the cover glass of Yama-Cam are (BK7, BK7 and borosilicate glass), while the dewer window, the cold window, the $H$-band pass filter of PRIME-Cam are (Synthetic fused silica, BK7, BK7). But the total thickness of the elements is the same.

\begin{figure}
 \begin{minipage}[b]{0.45\linewidth}
  \centering
  \includegraphics[width=95mm]{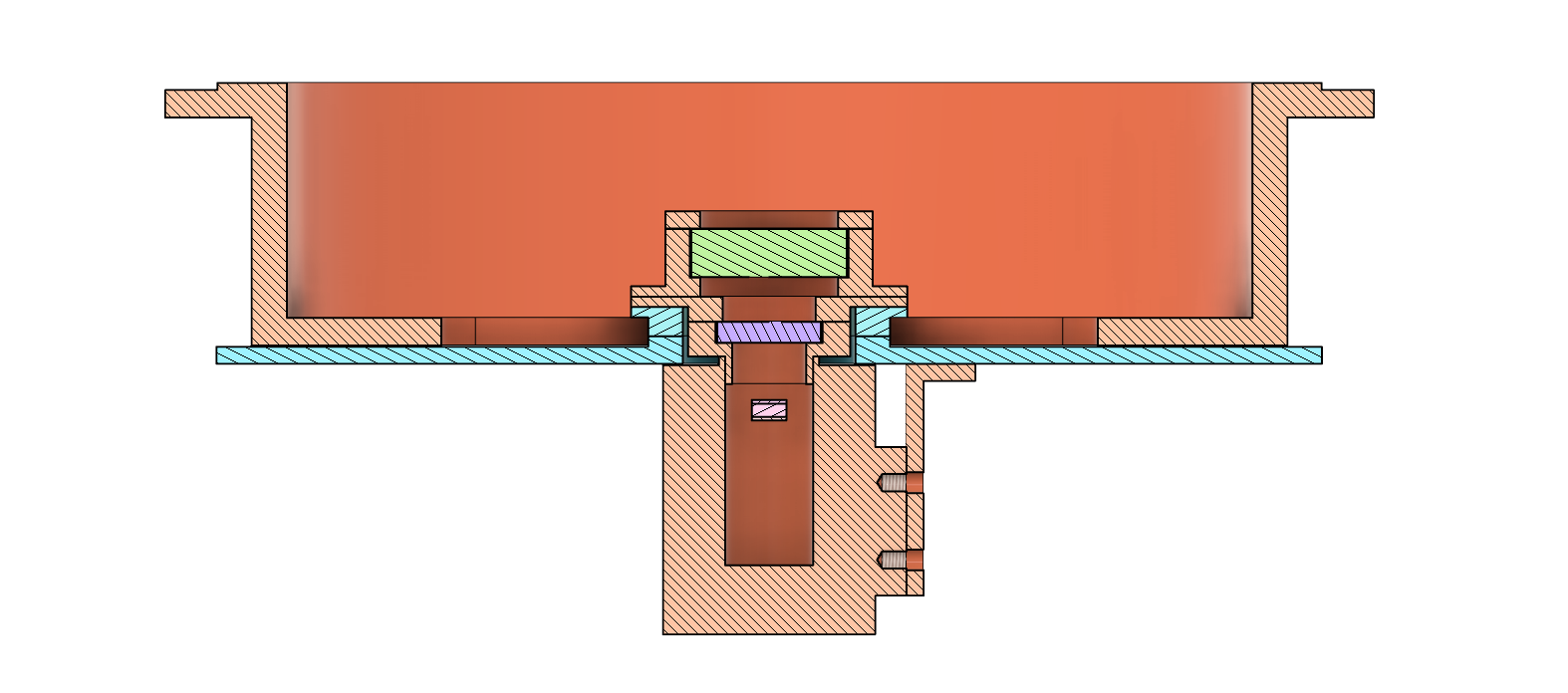}
  \subcaption{The cross-section of Yama-Cam.}
 \end{minipage}
 \begin{minipage}[b]{0.45\linewidth}
  \centering
  \includegraphics[width=60mm]{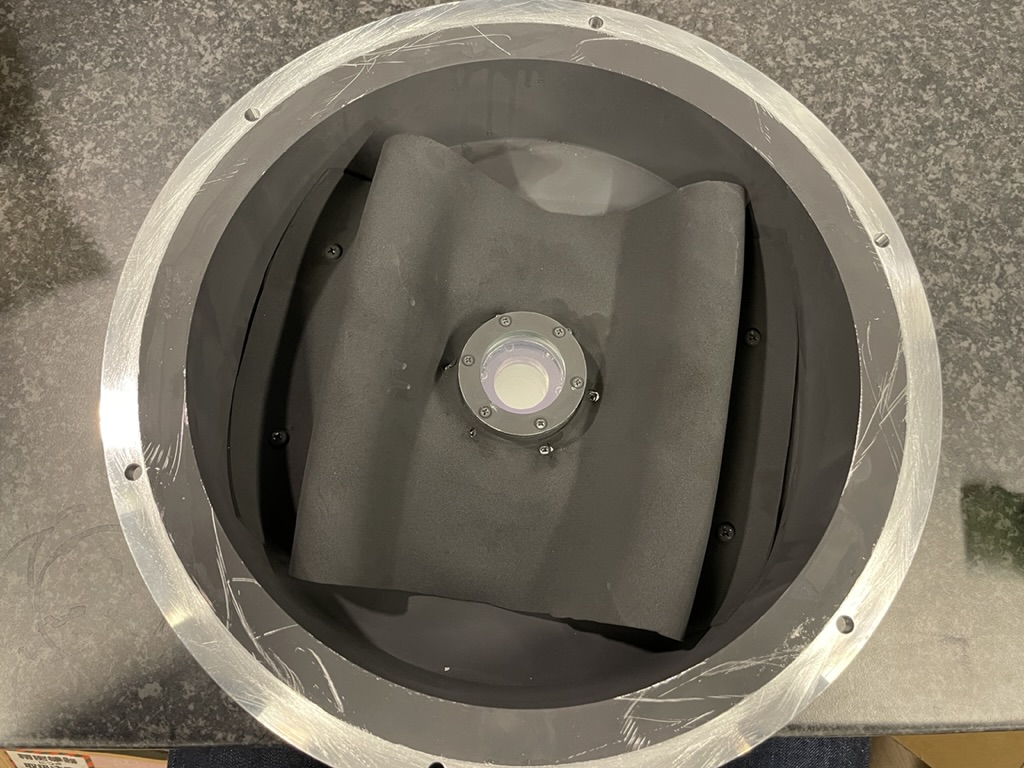}
  \subcaption{Picture of Yama-Cam.}
 \end{minipage}
 \caption{In the cross-section of Yama-Cam, the green object is the BK7 glass plate. The purple object is the $H$-band pass filter. The pink unit is the IMX990 section. The light blue object is the slide system.}
 \label{fig:YamaCamCross-section}
\end{figure}

One of the most important features is that the parts of filters and the detector can move along with the focal plane using a slide (slide system), which enables taking images across the wide FOV of PRIME. The pixel scale of Yama-Cam is $0.256"/\rm pix$ and the FOV is $6.45 \times 10^{-3}\, \rm{deg}^2$.
This FOV is much narrower than that of PRIME-Cam ($1.45\,\rm{deg}^2$), but the sliding system allows Yama-Cam to cover the wide FOV of PRIME-Cam. In other words, Yama-Cam can do optical alignment and observe in the same FOV of PRIME-Cam. 
Figure \ref{fig:YamaCamCross-section} shows a cross-section and a picture of Yama-Cam.

\subsubsection{Pseudo Hartmann Test}
A Hartmann test evaluates the optical performance by 
measuring the spot locations at the focal plane that 
are identified by connecting the corresponding spots 
in the intra and extra defocused images.
These two defocused images through the Hartmann plate with enough small holes are usually taken for
the identical optical system with the only difference of the intra and extra defocus. 
However, the focus adjustment mechanism of the PRIME telescope changes the position of the whole lens unit (PFU) with the mounted camera together along the optical axis.
Therefore, inspected optics are not identical between 
the intra and extra defocused images. Hence, we call 
our Hartmann test a pseudo Hartmann test.
The Hartmann constant derived from the pseudo Hartmann test for the PRIME optical system is consistent with that derived from the true Hartmann test. We confirmed this by the simulation. The difference is 0.002" and 0.003" for the center and corner of the FOV, respectively, and these are negligible compared to the typical uncertainty of the Hartmann constant measurement, $\sim0.02"$.

\begin{figure}
    \centering
    \includegraphics[width=140mm]{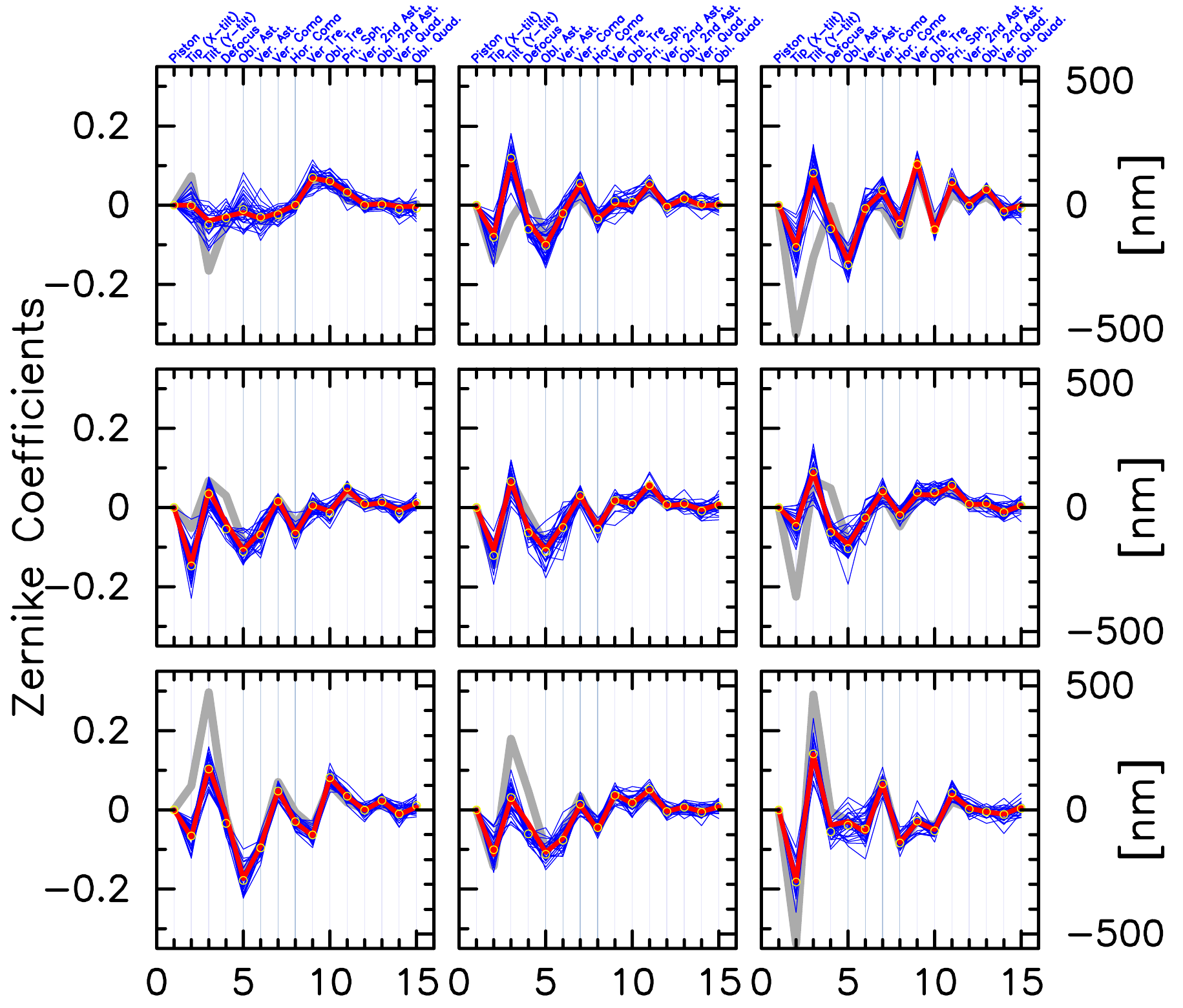}
    \caption{The standard Zernike coefficients (up to the 15 terms) at the nine different FOV positions. The gray lines show the coefficients for the nominal optics derived from ZEMAX. The red lines are the median values from the Pseudo Hartmann simulation (the blue lines) where Gaussian noise is considered for the spot detection assuming the seeing effect. Yellow circles show the estimated Zernike coefficients without the seeing effect. At the FOV center, the Zernike coefficients from ZEMAX and the Pseudo Harmann simulation are almost perfectly consistent, but at the FOV corners, the coefficients of tip, tilt and defocus are different.}
    \label{fig:pseudoH_zernike}
\end{figure}

The pseudo Hartmann test shows some amounts of differences in the derived Zernike terms obtained from ZEMAX simulations, especially for the outer FOV as shown in Figure \ref{fig:pseudoH_zernike}. 
This effect, however, is negligible for the optical alignment for the below reasons.
First, the major aberrations we correct by the decenter and CNR are coma and astigmatism and these Zernike coefficients are well estimated by the pseudo Hartmann test. 
Second, the difference in the defocus term is small enough compared to the tolerance.
Third, the tip and tilt can be probed by measuring the focal plane which corresponds to the minimum confusion circle.

By using ZEMAX, we generate a lot of pseudo Hartmann simulations by changing the decenter and CNR of the lens unit and L2 tip/tilt for the nine positions of the FOV, which are shown in Figure \ref{fig:LinearitySimulations}.
In the ZEMAX model we use for this simulation, the interferometer measurement information of PFU is added to the ZEMAX model used in Subsection \ref{sec_IFEF}.
This interferometer measurement information includes manufacturing errors and misalignment of lenses in PFU, and they produce higher order aberrations such as trefoil and quadratic astigmatism in the pseudo Hartmann simulation as in actual observations. 
We build linear models that relate the amount of the changes in each optical piece and Zernike coefficients. Table \ref{table:LinearitySimulation} summarized the amount of optics misalignment for the simulation.

\begin{table}
\caption{The amount of misalignment used to build models and corresponding aberrations.}
\begin{center}
\begin{tabular}{rccc}
\hline\hline
Misalignment & Amount of change & unit & Aberration \\ \hline 
Decenter X & -0.5,\,-0.2,\,-0.1,\,0.1,\,0.2,\,0.5 & mm & Tip,\,Horizontal Coma\\ 
Y & -0.5,\,-0.2,\,-0.1,\,0.1,\,0.2,\,0.5 & mm & Tilt,\,Vertical Coma  \\ 
CNR Tilt Rx & -360,\,-180,\,180,\,360 & as & Oblique Astigmatism,\,Vertical Astigmatism\\ 
Ry & -360,\,-180,\,180,\,360 & as & Oblique Astigmatism,\,Vertical Astigmatism\\ 
L2 Tilt Rx & -360,\,-180,\,180,\,360 & as & Tilt,\,Vertical Coma\\ 
Ry & -360,\,-180,\,180,\,360 & as & Tip,\,Horizontal Coma\\ \hline
\end{tabular}
\label{table:LinearitySimulation}
\end{center}
\end{table}

The Zernike coefficients $\mathbf a$ are derived by minimizing $\Delta$,
\begin{equation}
    \Delta = \sum_{m}[\sum_{n}\mathbf Z_{n,m} \mathbf a_{n} - \mathbf B_{m}]^2
\end{equation}
where $\mathbf Z_{n,m}$ is the $n$th differential Zernike polynomial term at the $m$th spot, $\mathbf a_n$ is the $n$th Zernike coefficient, $\mathbf B_{m}$ is the measured slope of the wavefront surface derived from the dislocation of the $m$th Hartmann spot.
The solution of the least square can be written as follows,
\begin{equation}
    \mathbf a = (\mathbf Z^{T}\mathbf Z)^{-1}\mathbf Z^{T}\mathbf{B}
\end{equation}
as $\mathbf{Z}^{T}\mathbf Z$ is a regular matrix.

The sensitivity of this method to PFU decenter, PFU CNR and L2 tip/tilt are simulated by using simulation Hartmann images under the average seeing condition at the Sutherland observatory. Even under the typical seeing, we found that several times iterations are enough to converge as indicated in Figure \ref{fig:pseudoH_zernike}, and the sensitivity to PFU decenter　is around ten $\mu$m, and PFU CNR and L2 tip/tilt are both within a few tens of arcseconds.
Since we confirmed that the models can be linearly interpolated and extrapolated with regard to FOV, we can use Hartmann spots at any location on the FOV when we use the PRIME-Cam.

\section{Result of the Optical Alignment Test}
\label{sec_res}
The PRIME telescope construction started at the beginning of July 2022. At the end of the construction phase, we conducted the optical alignment test with the method described in this paper.

In all of the alignment methods, basically, we adjusted PFU decenter and PFU tip/tilt by turning the adjustment screws manually. The adjustment was made with L2 tip/tilt, which is adjustable by the actuators, only when the rotation angle of the adjustment screws corresponding to the amount of correction of PFU decenter or PFU tip/tilt was too small.

We describe the result of the optical alignment test in Japan and that in Sutherland below. 

\subsection{Result of the Optical Alignment Test in Japan}
We tested the construction of the PRIME telescope at Nishimura Co., Ltd in Shiga, Japan in 2020.
We also tested the rough alignment by the laser tracker and the fine alignment in $z$-band by the IFEF image technique. 

 \subsubsection{Result of the Preliminary Alignment by Laser Tracker in Japan}
 First, we did the initial alignment by FARO. Considering the accuracy of the laser tracker and this method, the goal of this method was set to achieve axes decenter $<50\,\mu$m and tip/tilt $<10"$. Note that even if this goal is satisfied, this method alone does not guarantee that the operational tolerance is met, since there is systematic uncertainty in the models to estimate the optical axis of PFU and mirror from the reference points for the SMR measurements as written in Subsection \ref{sec_faro}

 We iterated to measure the reference points and adjust the optical axes several times and achieved PFU decenter $38.6\,\mu$m and PFU tip/tilt $2.0"$.

 \subsubsection{Result of the Fine Alignment in $z$-band by IFEF Image Technique in Japan}
 We set the goal value at PFU decenter $<280\,\mu$m and PFU CNR $<40"$. Note that the goal value of PFU CNR is derived from the conversion of PFU tip/tilt value of the operational tolerance.

 We iterated to do this method and achieved PFU decenter of $\simeq100\,\mu$m and PFU CNR of $\simeq40"$. The average Hartmann constant for the FOV of the PRIME-Cam was $0.41$ at $z$-band.

\subsection{Result of the Optical Alignment Test in Sutherland}
We confirmed that the preliminary alignment by laser tracker and the fine alignment in $z$-band by IFEF image technique can improve optics in Japan. 
In Sutherland, we conducted two optical alignment steps we tested in Japan and the fine alignment by Hartmann test.

\subsubsection{Result of the Preliminary Alignment by Laser Tracker in Sutherland}
\label{sec_res_las}
We followed the same procedure as we did in Japan. The goal values are the same as in Japan. As a result, we achieved PFU decenter of $46.4\,\mu$m and PFU tip/tilt of $12.4"$. Although PFU tip/tilt was over the goal a little, we finished this method, since the rotation angle of the bolt required to adjust PFU tip/tilt was as small as 6.6 deg. 

We measured the Hartmann constant to confirm the result of the initial alignment by FARO, just after the installation of the $z$-band camera.
We achieved the Hartmann constant of $0.57"$. We also mounted Yama-Cam to check the throughput with the $H$-band filter by using a bright star. Thanks to the rough alignment by FARO, we were able to achieve the $H$-band first light as well.

\subsubsection{Result of the Fine Alignment in $z$-band by IFEF Image Technique in Sutherland}
\label{sec_res_IFEF}
After the first light in $z$- and $H$-band, we immediately moved to the fine alignment in $z$-band by the IFEF image technique. We conducted this method to satisfy the same goal values of the test in Japan, however, there were software problems and the amount of misalignment did not converge within the goal values. The Hartmann constant was 0.92" for the FOV of PRIME-Cam after the last alignment of this method. We could not improve the PRIME optics using this method at Sutherland. 

As we have limited time for the optical alignment test, we decided to do the fine alignment by the Hartmann test with $z$-band\footnote{We made new model for $z$-band by using ZEMAX immediately.}. The result of the fine alignment by the Hartmann test with $z$-band is written in Subsection \ref{sec_res_Har}.

\subsubsection{Result of the Fine Alignment by Hartmann Test in Sutherland}
\label{sec_res_Har}
We used the Hartmann constant as the goal value of this method. In $z$-band, we set the goal value at Hartmann constant $\simeq$ 0.41" which we had achieved in Japan. We iterated the fine alignment by the Hartmann test with $z$-band and we achieved 0.44" of Hartmann constant. As a result, we confirmed that this method improves the optics. Then, we immediately replaced the $z$-band camera with Yama-Cam for the fine alignment in $H$-band by the Hartmann test. The Hartmann constant in $H$-band was 0.509" after the $H$-band camera was installed.

The goal value was set at the average Hartmann constant for the inner FOV $<$ 0.314" which was calculated to meet the operational tolerance. We conducted the fine alignment in $H$-band by the Hartmann test repeatedly, and the smallest Hartmann constant we achieved is 0.262" for the center of FOV and 0.295" for the inner FOV (which is roughly a circle with a radius of $0.555$ deg.). After we achieved the smallest Hartmann constant, additional work on the telescope made the optics a little worse. Therefore, we did this method again. As a result, we achieved 0.322" of the average Hartmann constant for the inner FOV of the PRIME-Cam. This is slightly larger than the expected value of 0.314", which is estimated with the nominal optical design plus the operational tolerance. However, the nominal design does not include all the systematics such as uncertainty in the primary mirror surface, rotator plate mounting on the PFU, camera mounting on the rotator plate, detector position, etc. Therefore, we consider the PRIME telescope after the alignment does satisfy the operational tolerance. Hence our alignment method is good enough to adjust the PFU to the primary mirror.

\section{Conclusion}
\label{sec_con}
The optical alignment method of the PRIME telescope is described in this paper. It consists of three steps; (1) preliminary alignment by using a laser tracker, (2) fine alignment by using IFEF image analysis, and (3) fine alignment by the pseudo Hartmann test.
The first two steps were demonstrated to work well in the optical alignment test in Japan in 2020. 
To optimize the optics in $H$-band and complement the second step where the $z$-band camera is used, we developed a new $H$-band camera and the third method.
We apply the methods in practice to conduct the optical alignment at the installation of the PRIME telescope. The results of the optical alignment methods are summarized in Table \ref{tab:results}.

\begin{table}
    \caption{The Comparison of Hartmann constant before and after each optical alignment. All Hartmann constants in this table are averages within the FOV of the PRIME-Cam.}
    \centering
    \begin{tabular}{clll} \hline
         Method & Sensitivity & \begin{tabular}{l}
             Hartmann const  \\
             before alignment 
         \end{tabular} & \begin{tabular}{l}
              Hartmann const \\
              after alignment
         \end{tabular}  \\ \hline \hline
         \begin{tabular}{c}
              Preliminary Alignment \\
              by Laser Tracker
         \end{tabular} & \begin{tabular}{l}
              $< 100\,\mu m$ (dec)  \\
              $<$ several tens as (tip/tilt)
         \end{tabular} & \begin{tabular}{c}
              -
         \end{tabular} & \begin{tabular}{l}
              0.57" ($z$)
         \end{tabular} \\ \hline
         \begin{tabular}{c}
              Fine Alignment \\
              by IFEF Image Teqnique
         \end{tabular} & \begin{tabular}{l}
              $<$ several tens\,$\mu m$ (dec)  \\
              $<$ a few tens as (CNR)
         \end{tabular} & \begin{tabular}{l}
              - (Japan)\\
              0.57" ($z$)
         \end{tabular} & \begin{tabular}{l}
              0.41" ($z$, Japan) \\
              0.92" ($z$)
         \end{tabular} \\ \hline
         \begin{tabular}{c}
              Fine Alignment in $z$-band  \\
              by Hartmann Test
         \end{tabular} & \begin{tabular}{l}
              $\sim$ ten\,$\mu m$ (dec)  \\
              $<$ a few tens as (CNR)
         \end{tabular} & \begin{tabular}{l}
              0.92" ($z$)
         \end{tabular} & \begin{tabular}{l}
              0.44" ($z$)\\
              0.509" ($H$)
         \end{tabular} \\ \hline
         \begin{tabular}{c}
              Fine Alignment in $H$-band  \\
              by Hartmann Test
         \end{tabular} & \begin{tabular}{l}
              $\sim$ ten\,$\mu m$ (dec)  \\
              $<$ a few tens as (CNR)
         \end{tabular} & \begin{tabular}{l}
              0.509" ($H$)
         \end{tabular} & \begin{tabular}{l}
              0.295" ($H$, best)
         \end{tabular}\\ \hline
    \end{tabular}
    \label{tab:results}
\end{table}

We achieved the average Hartmann constant of $0.295"$ for the inner FOV with Yama-Cam, the $H$-band test camera.
The final optical alignment with PRIME-Cam, the primary NIR imaging instrument of the PRIME telescope will be conducted in early 2023. The Galactic center time domain survey is expected to start in 2023. 

\section*{Acknowledgments}
We thank Nishimura Co.,Ltd for letting us use their laboratory and a FARO laser tracker for the optical alignment test.
The PRIME project is supported by JSPS KAKENHI Grant Number 16H06287, 19KK0082, 20H04754, 22H00153 and JPJSCCA20210003.
These results are based on data obtained from PRIME at the South African
Astronomical Observatory (SAAO), Sutherland, South Africa.

\bibliographystyle{ws-jai}
\bibliography{main}

\end{document}